\documentclass[11pt,cite,epsfig,psfrag]{article}
\usepackage[utf8]{inputenc}
\usepackage{placeins}
\usepackage{geometry}
 \usepackage{graphicx}
 \usepackage{subfig}
  \usepackage{amsmath} 
  \graphicspath{{images33/}}
  \usepackage[usenames, dvipsnames]{color}
\usepackage{wrapfig}
\usepackage{boxedminipage}

\usepackage{fullpage}
\usepackage{amsmath}
\usepackage{amssymb}
\usepackage{graphicx}
\usepackage{mathrsfs}
\usepackage{wrapfig}
\usepackage{boxedminipage}
\usepackage{epsfig}
\usepackage{setspace}
\usepackage{float}
\usepackage{hyperref}
\usepackage{enumerate}
\usepackage{cite}
\usepackage[font=footnotesize,labelfont=rm]{caption}

\usepackage[numbers,sort&compress]{natbib}

\def\be{\begin{equation}}
\def\ee{\end{equation}}
\def\bea{\begin{eqnarray}}
\def\eea{\end{eqnarray}}

\numberwithin{equation}{section}

 \newcommand{\RN}[1]{%
   \textup{\uppercase\expandafter{\romannumeral#1}}%
 }
\begin{document}

\thispagestyle{empty}

\vskip 2cm

\begin{center}
{\Large \bf A Note on Gauss-Bonnet Black Holes at Criticality}
\end{center}

\vskip .2cm

\vskip 1.2cm

\centerline{ \bf Chandrasekhar Bhamidipati\footnote{chandrasekhar@iitbbs.ac.in} and 
 Pavan Kumar Yerra \footnote{pk11@iitbbs.ac.in}
}

\vskip 7mm 
\begin{center}{ School of Basic Sciences\\ 
Indian Institute of Technology Bhubaneswar \\ Bhubaneswar 751013, India}
\end{center}

\vskip 1.2cm
\vskip 1.2cm
\centerline{\bf Abstract}
\noindent
Within the extended thermodynamics, we give a comparative study of 
critical heat engines for Gauss-Bonnet and charged black holes in AdS in five dimensions, in the limit of large Gauss-Bonnet parameter $\alpha$ and charge $q$, respectively. We show that the approach of efficiency of heat engines to Carnot limit in Gauss-Bonnet black holes is higher(lower) than charged black holes when corresponding parameters are small(large). 


\newpage
\setcounter{footnote}{0}
\noindent

\baselineskip 15pt

\section{Introduction}

Recently, the physics of charged black holes in AdS~\cite{Chamblin:1999tk,PhysRevD.60.104026} in the neighbourhood of a second order phase transition has been formulated in a novel way~\cite{Johnson:2017hxu,Johnson:2017asf}. At this critical point, there is a scaling symmetry where the thermodynamic quantities scale with respect to charge $q$, i.e., Entropy $S\sim q^2$, Pressure $p\sim q^{-2}$, and Temperature $T\sim q^{-1}$. Interestingly, it has been shown that geometry of black hole near the critical point yields a fully decoupled Rindler space-time in the double limit of nearing the horizon, while at the same time keep the charge of black hole large. These results might have profound implications for holography constructions where Rindler space appears in the decoupling limit. This novel approach near the critical point might shed further light on holography in Rindler spacetime. The physics of the geometry near the critical point itself is quite interesting from the gravity side.\\

\noindent
Study of the critical region of black holes, has been facilitated by the existence of an extended thermodynamic description of charged black holes in AdS, which shows a phase structure that includes a line of first order phase
transitions ending in a second order transition point~\cite{Chamblin:1999tk,Kubiznak:2012wp,Gunasekaran:2012dq,Cai:2013qga,Kubiznak:2016qmn}. In this context, apart from the Hawking-Page transition (connecting black holes in AdS to large N gauge theories at finite temperature), Van der Waals transition has captured the attention recently. A holographic interpretation for the later transition was proposed in~\cite{Johnson:2014yja}, where it is interpreted not as a thermodynamical transition but, instead, as a transition in the space of field theories (labeled by N, the number of colors in the gauge theory). Thus, varying the cosmological constant in the bulk corresponds to perturbing the dual CFT, triggering a renormalization group flow. This flow is captured in the bulk by Holographic heat engines with black holes as working substances~\cite{Johnson:2014yja}. Various aspects of this correspondence are being actively studied both from the gravity point of view as well as for potential applications to the dual gauge theory side~\cite{Dolan:2011xt,Dolan:2010ha,Kastor:2009wy,Caldarelli:1999xj,Sinamuli:2017rhp,Karch:2015rpa,Kubiznak:2014zwa,Johnson:2014yja,Johnson:2015ekr,Johnson:2015fva,Belhaj:2015hha,Bhamidipati:2016gel,Chakraborty:2016ssb,Hennigar:2017apu,Johnson:2017ood,Setare2015,Caceres2015,Mo2017,Liu:2017baz,Wei:2016hkm,Sadeghi:2015ksa,Zhang:2016wek,Kubiznak:2016qmn,Sadeghi:2016xal}. Furthermore, a holographic heat engine defined at this critical point has the special property that its efficiency approaches that of Carnot engine at finite power\footnote{ See~\cite{PhysRevLett.106.230602,PhysRevLett.111.050601,PhysRevLett.115.090601,PhysRevX.5.031019,PhysRevLett.114.050601,power_of_a_critical_heat,Koning2016,PhysRevLett.117.190601,eff_vs_speed}, for recent discussions on approaching Carnot limit at finite power in thermodynamics and statistical mechanics literature.}, as the charge parameter $q \rightarrow \infty$~\cite{Johnson:2017hxu}. \\

\noindent
Intrigued by the above developments, in this note, we study holographic heat engines 
for Gauss-Bonnet (GB) black holes in AdS, whose phase structure closely resembles that of charged black holes, where the role of the charge parameter $q$ is played by the GB parameter $\alpha$. We analyze properties of heat engines at the critical point, following the methods proposed by Johnson in~\cite{Johnson:2017hxu} and compare the efficiencies as a function of $\alpha$ and $q$. The motivations are as follows. Higher derivative curvature terms such as Gauss-Bonnet terms occur in many occasions, such as in  the semiclassical quantum gravity and in the effective low-energy effective action of superstring theories. In the latter case, according to the AdS/CFT correspondence~\cite{Aharony:1999ti,Maldacena:1997re}, these terms can be viewed as the corrections of large $N$  expansion of boundary CFTs in the strong coupling limit. It is also known that such corrections have interesting consequences to viscosity to entropy ratio~\cite{PhysRevD.77.126006}. In this spirit, 
corrections to the efficiency of heat engines (with charged black holes as working substances) coming from GB terms were considered in detail in~\cite{Johnson:2015ekr}. It was noted that the efficiency of the engine depends on which parameters of the engine are held fixed as $\alpha_{GB}$ is changed. It could increase or decrease, depending on the scheme used. Here, our interest is in the critical region.\\


\noindent
Charge neutral AdS Schwarzschild black hole is not very useful as a heat engine as the specific heat at constant pressure is always negative for small black hole. However, in the case of charged AdS black holes (with or with out the Gauss-Bonnet term) heat engines can be defined with holographically dual field interpretation. In the case of neutral black holes, the presence of a GB parameter, however, allows for phase transitions and PV criticality akin to charged black holes in AdS. This has been noted in~\cite{Cai:2013qga,Mo:2014qsa}, where it was pointed out that GB parameter mimics the charge. Based on the PV critical behavior it is possible to define a heat engine for charge neutral Gauss-Bonnet black holes as working substances. The key difference from~\cite{Johnson:2015ekr}, is that there the GB coupling was a parameter and the working substance was a charged black holes. In the present context, the equation of state one uses corresponds to neutral GB black holes, which are themselves the working substances. We also compare, how the approach of efficiency of engines to the Carnot limit is in GB and charged black holes, coming from $q$ as well as $\alpha$~\cite{Johnson:2017hxu}. This is important because, as compared to charged black holes, the parameter $\alpha$ takes care of the corresponding next to leading order corrections in the large $N$ limit in the gauge theory. 

 \vspace{0.7cm} \noindent 
Consider the action for D-dimensional Einstein theory with a Gauss--Bonnet term and a cosmological constant $\Lambda$ as~\cite{Cai:2013qga,Cai:2001dz,Johnson:2015ekr}:
\begin{equation}
I=\frac{1}{16\pi}\int\! d^Dx \sqrt{-g}\left[R-2\Lambda +\alpha_{\rm GB}(R_{\gamma\delta\mu\nu}R^{\gamma\delta\mu\nu}-4R_{\mu\nu}R^{\mu\nu}+R^2)\right]\ ,
\label{eq:gauss-bonnet-action}
\end{equation}
where  the Gauss--Bonnet parameter $\alpha_{\rm GB}$ has dimensions of ${\rm (length)}^2$ and the cosmological constant is
\begin{equation}
\Lambda=-\frac{(D-1)(D-2)}{2l^2}. 
\label{eq:cosmocon}
\end{equation} 
The  action admits  a   static black hole solution  with the metric:
\begin{equation}
ds^2 = -Y( r)dt^2
+ {dr^2\over Y(r)} + r^2 d\Omega^2_{D-2} 
\label{eq:staticform}
\end{equation}
where
\begin{equation}
Y(r)=1+\frac{r^2}{2\alpha}\left(1-\sqrt{1+\frac{4\alpha m}{r^{D-1}}-\frac{4\alpha}{l^2}}\right) \ ,
\label{eq:why}
\end{equation}

Here, $d\Omega^2_{D-2}$ is the metric on a round $D-2$ sphere with volume $\omega_{D-2}$ and  $\alpha=(D-3)(D-4)\alpha_{\rm GB}$. At $r=r_+$, is the largest positive real root of $Y(r)$. The mass of the solution is given by~\cite{Cai:2001dz,Cai:2013qga,Johnson:2015ekr}:
\begin{equation}
M=\frac{(D-2)\omega_{D-2}}{16\pi} m \, .
\label{eq:paramters}
\end{equation}
In order to have a well defined vacuum solution (with $m=0$),  for a given value of $l$ (and hence $\Lambda$) $\alpha$ cannot be arbitrary~\cite{PhysRevLett.55.2656}, but in fact must be constrained by $0\leq {4\alpha}/{l^2}\leq 1$. For later use we can write this in terms of the pressure ( using $ p=-\Lambda/8\pi$) as:

\begin{equation}
0\leq \alpha \leq \alpha_*\ , \quad {\rm where}\quad  \alpha_*={(D-1)(D-2)}/{64\pi p} \ .
\label{eq:constrain-alpha}
\end{equation}

The  horizon radius $r_+$ of the black hole is set by the largest  root of $Y(r_+)=0$, which gives us an equation for  $M$, 
\begin{equation}
M=\frac{(D-2)\omega_{D-2}}{16\pi}\left(\alpha r_+^{D-5}+r_+^{D-3}+16\pi p \frac{r_+^{D-1}}{(D-1)(D-2)}\right)\ ,
\end{equation}
where we have replaced $l$ by $p$ using $p=-\Lambda/8\pi$ and equation~(\ref{eq:cosmocon}). 
The temperature comes from the first derivative of $Y$ at the horizon, in the usual way:
\begin{equation}
T=\frac{Y^\prime(r_+)}{4\pi}=\frac{1}{4\pi r_+(r_+^2+2\alpha)}\left(\frac{16\pi p r_+^4}{(D-2)}+(D-3)r_+^2+(D-5)\alpha\right)\ ,
\label{eq:GBeqnofstate}
\end{equation}
The function $M$ defines our enthalpy $H(p,S)$, from which the entropy  can be computed as:
\begin{equation}
S=\int_0^{r_+}\frac{1}{T}\left.\frac{\partial M}{\partial r}\right|_{p} dr=\frac{\omega_{D-2}}{4} r_+^{D-2}\left(1+\frac{2(D-2)}{(D-4)}\frac{\alpha}{r_+^2}\right)\ .
\end{equation}
and the thermodynamic volume is given by
\begin{equation}
V=\frac{\omega_{D-2}}{(D-1)} r_+^{D-1}\ .
\label{eq:volume}
\end{equation} 
 Holographic heat engine can be defined for extracting mechanical work from heat energy via the $pdV$ term present in the First  Law of extended black hole thermodynamics~\cite{Johnson:2014yja}, where, the working substance is a black hole solution of the gravity system. One starts by defining a cycle in state space where there is a net input heat flow $Q_H$, a net output heat flow $Q_C$, and a net output work W, such that $Q_H = W + Q_C$. The efficiency of such heat engines can be written in the usual way as $\eta=W/Q_H=1-Q_C/Q_H$. Formal computation of efficiency proceeds via the evaluation of $\int C_p dT$ along those isobars, where~$C_p$ is the specific heat at constant pressure or through an exact formula by evaluating the mass at all four corners as~\cite{Johnson:2015ekr,Johnson:2015fva,Johnson:2016pfa}: 
 \begin{equation}
 \eta = 1- \frac{M_3 - M_4}{M_2 - M_1} \, . \label{eq:efficiency-prototype} 
 \end{equation}
In this note, we are interested in the case neutral Gauss-Bonnet black holes as working substances for heat engines. Before proceeding to analyze the behavior of heat engines at criticality, we first present
few computations of efficiency of heat engines in neutral GB black holes. For black holes with charge and GB corrections, results were reported in ~\cite{Johnson:2015ekr}, but, there the working substance was a charged black holes, which provides the equation of state. In the present case, the Gauss-Bonnet black hole itself is the working substance, giving a new equation of state (\ref{eq:equationofstate}) and a priori, it is not clear how the efficiency of the heat engine should behave, if the charge parameter is set to zero.\\

\noindent
In $D=5$, the expressions for Mass $M$ and temperature $T$ for GB black holes read as~\cite{Cai:2013qga}:
\begin{equation}
M\equiv H=\frac{3\pi}{8}\Big(\alpha + r_+^2 + \frac{4\pi pr_+^4}{3}\Big) ,
\label{eq:enthalpy} 
\end{equation}
and
\begin{equation}
T=\frac{(2V)^{\frac{1}{4}}}{(2\pi)^{\frac{3}{2}}\Big(\frac{\sqrt{V}}{\pi} + \alpha\sqrt{2}\Big)}\Bigg(1+\frac{8p\sqrt{2V}}{3}\Bigg) .
\label{eq:temperature}
\end{equation}
An equivalent expression to eq.~(\ref{eq:temperature}) is the equation of state $p(V,T)$:
\begin{equation}
p=\frac{3}{8}\Bigg\{T\Big(\pi\sqrt{\frac{2}{V}}\Big)^{\frac{3}{2}}\Big(\frac{\sqrt{V}}{\pi} + \alpha\sqrt{2}\Big)-\frac{1}{\sqrt{2V}}\Bigg\} .
\label{eq:equationofstate}
\end{equation}
 \begin{figure}[h]
 	{\centering
 		\subfloat[]{\includegraphics[width=2.5in]{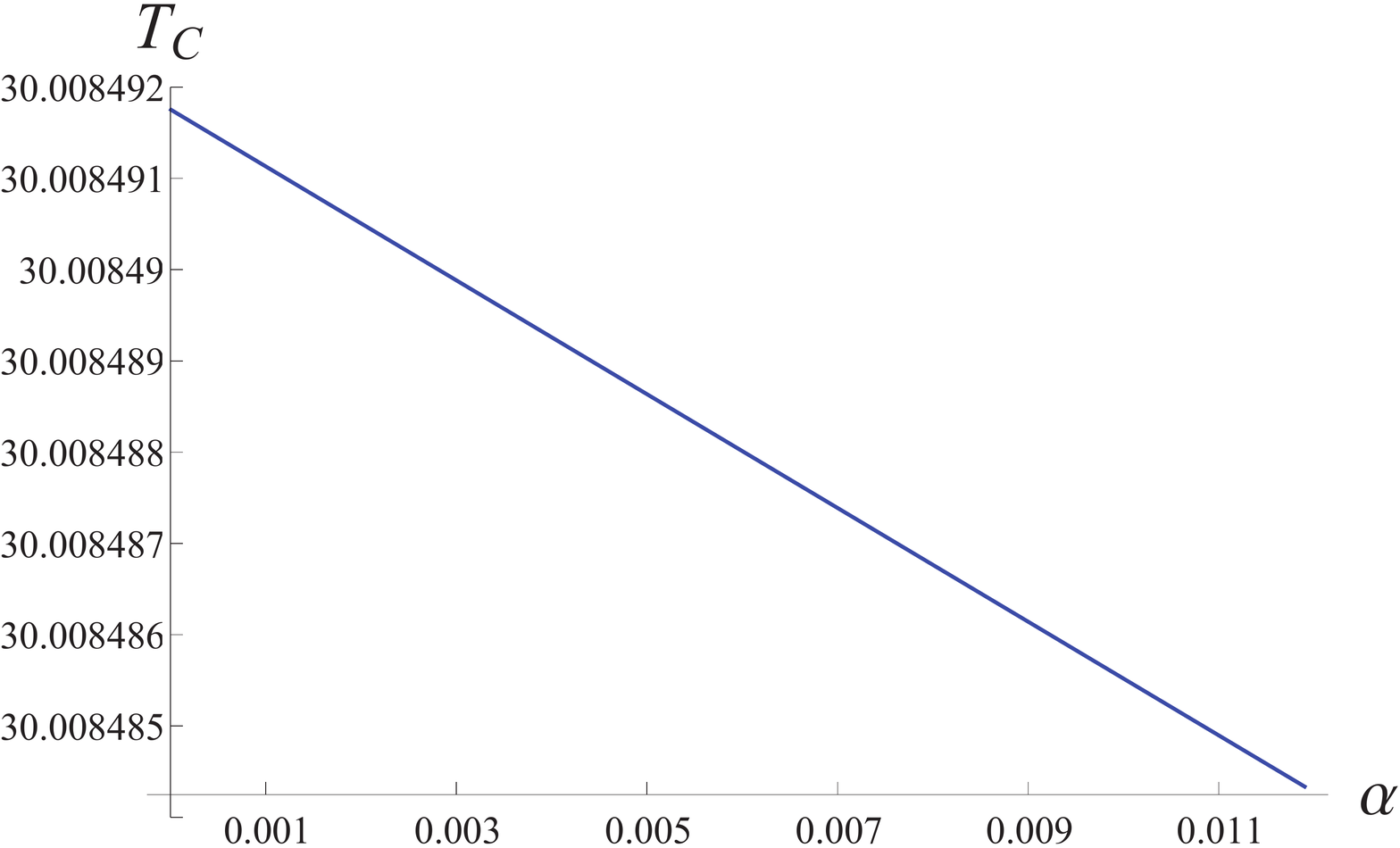} }\hspace{1.9cm}
 		\subfloat[]{\includegraphics[width=2.5in]{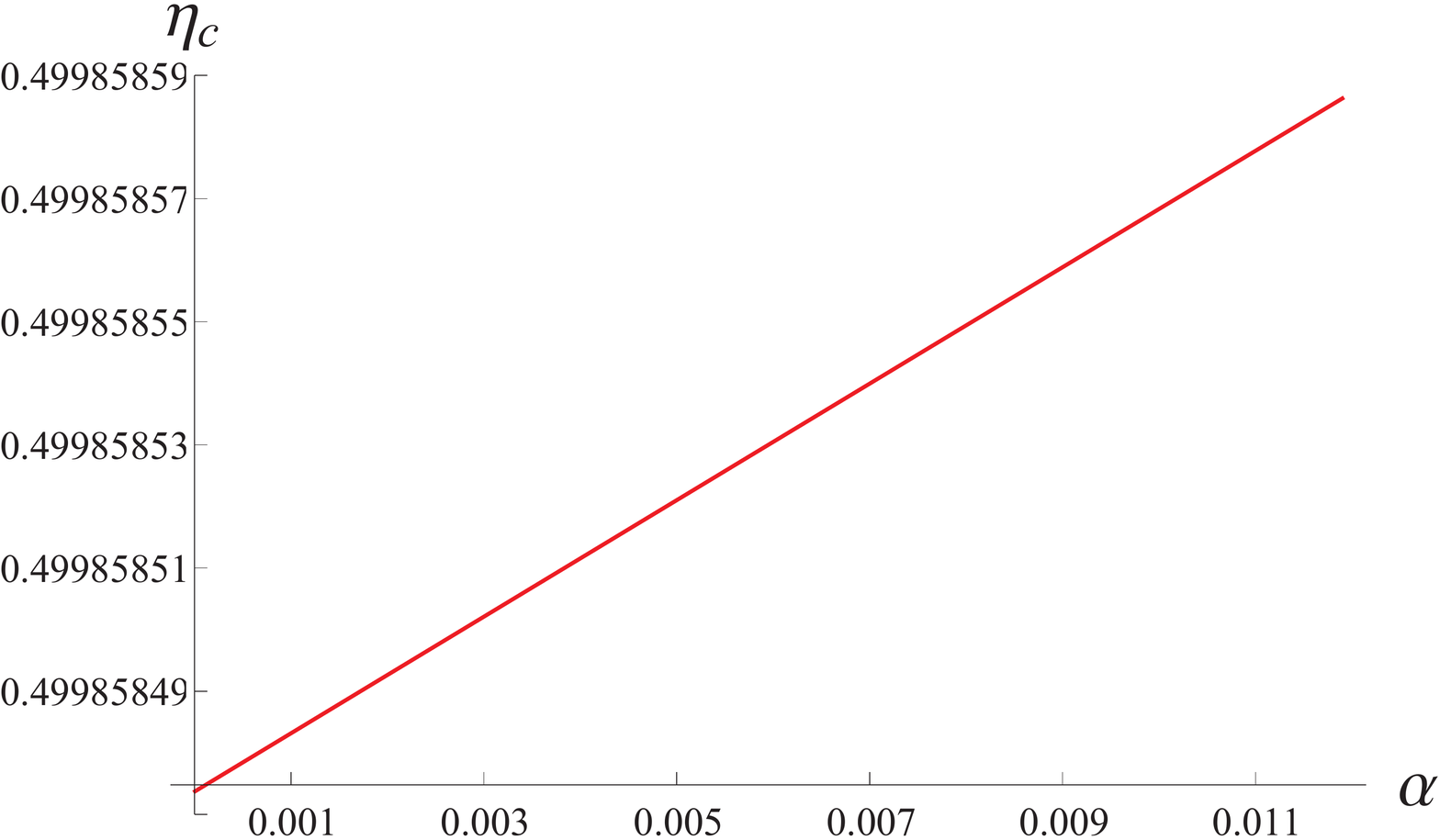} }
 		
 		\caption{\footnotesize In scheme 1, over the  physical range of $\alpha$ constrained  by the relation~(\ref{eq:constrain-alpha}), (a) The lowest temperature of the engine $T_C$ {\it vs} $\alpha$,  and (b)  The Carnot's efficiency $\eta_{\rm C}$ {\it vs} $\alpha$.   (Here, we have chosen the values $p_1=5, p_4=3, T_1=50$, and $T_2=60$.)}   \label{fig:scheme1_TC_and_itac}
 	}
 \end{figure}
 
  \begin{figure}[h]
  	{\centering
  		\subfloat[]{\includegraphics[width=2.5in]{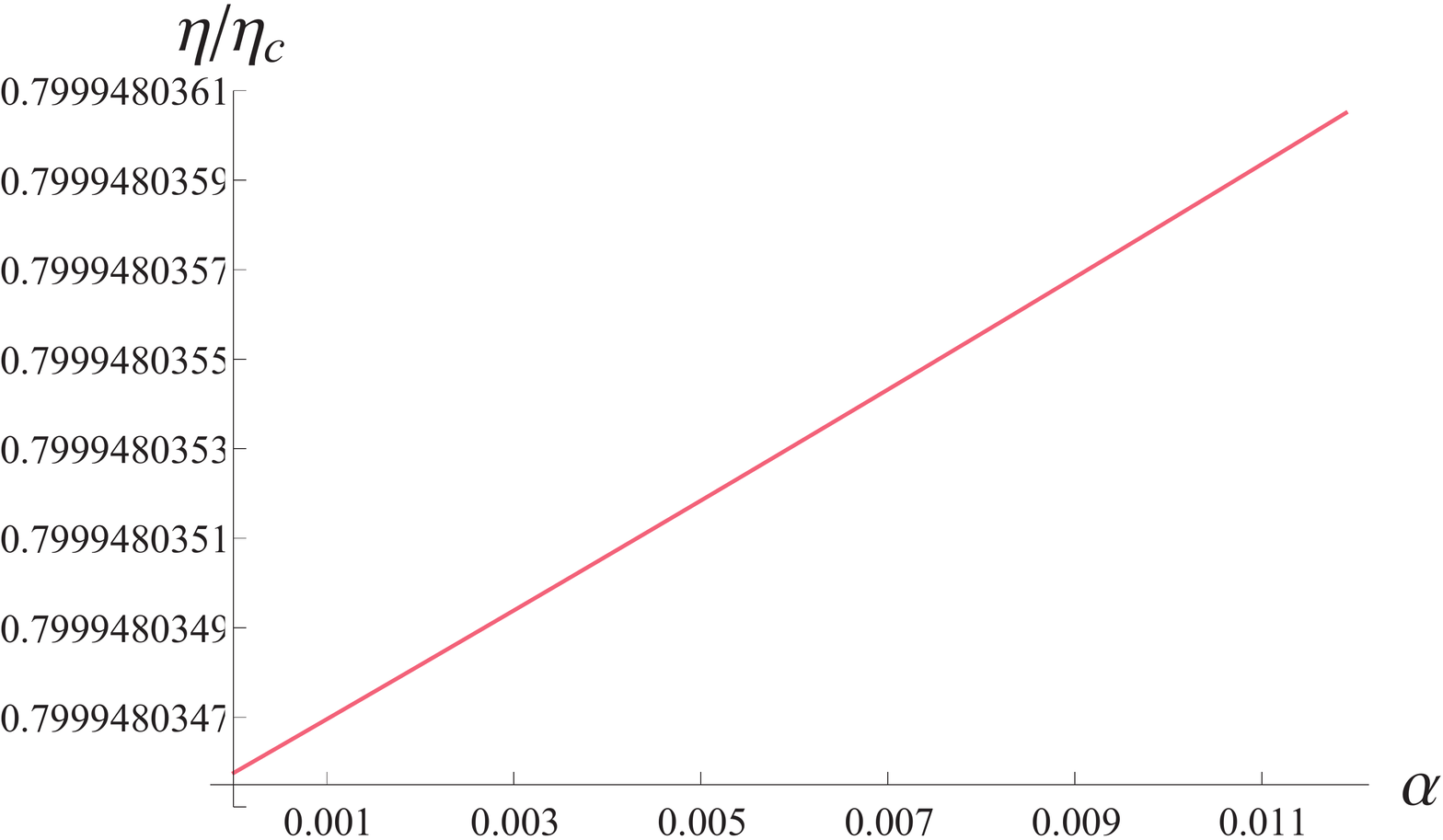} }\hspace{1.9cm}  	
  		\subfloat[]{\includegraphics[width=2.5in]{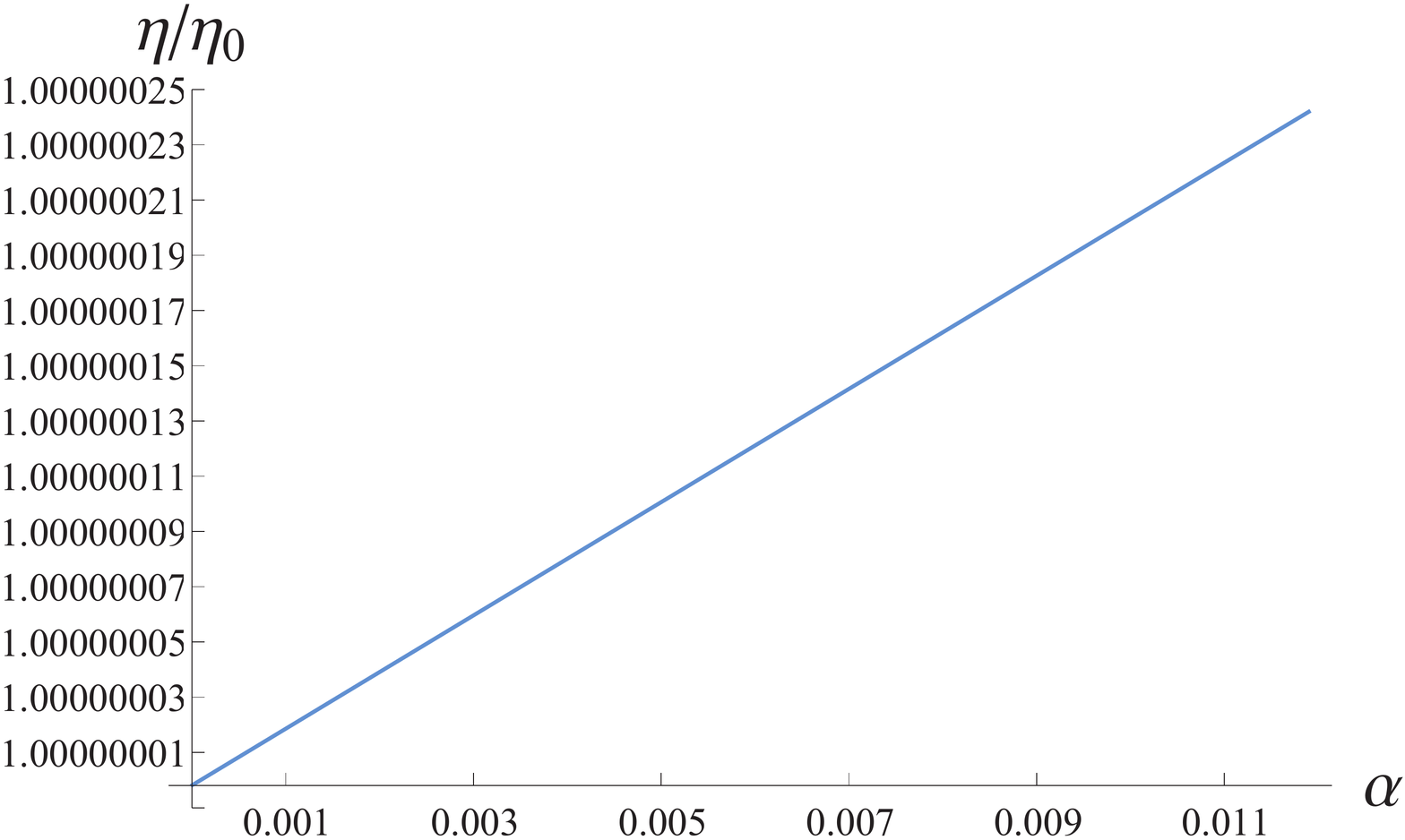} }
  		
  		\caption{\footnotesize  In scheme 1, over the  physical range of $\alpha$ constrained  by the relation~(\ref{eq:constrain-alpha}), (a) The ratio $\eta/\eta_{\rm C}$ {\it vs} $\alpha$, and  (b)  The ratio $\eta/\eta_0$  {\it vs} $\alpha$.   (See the  caption of fig (\ref{fig:scheme1_TC_and_itac}) for parameter values.)}   \label{fig:scheme1_compare}
  	} 
  \end{figure}

 \begin{figure}[h!]
 	{\centering
 		\includegraphics[width=2.8in]{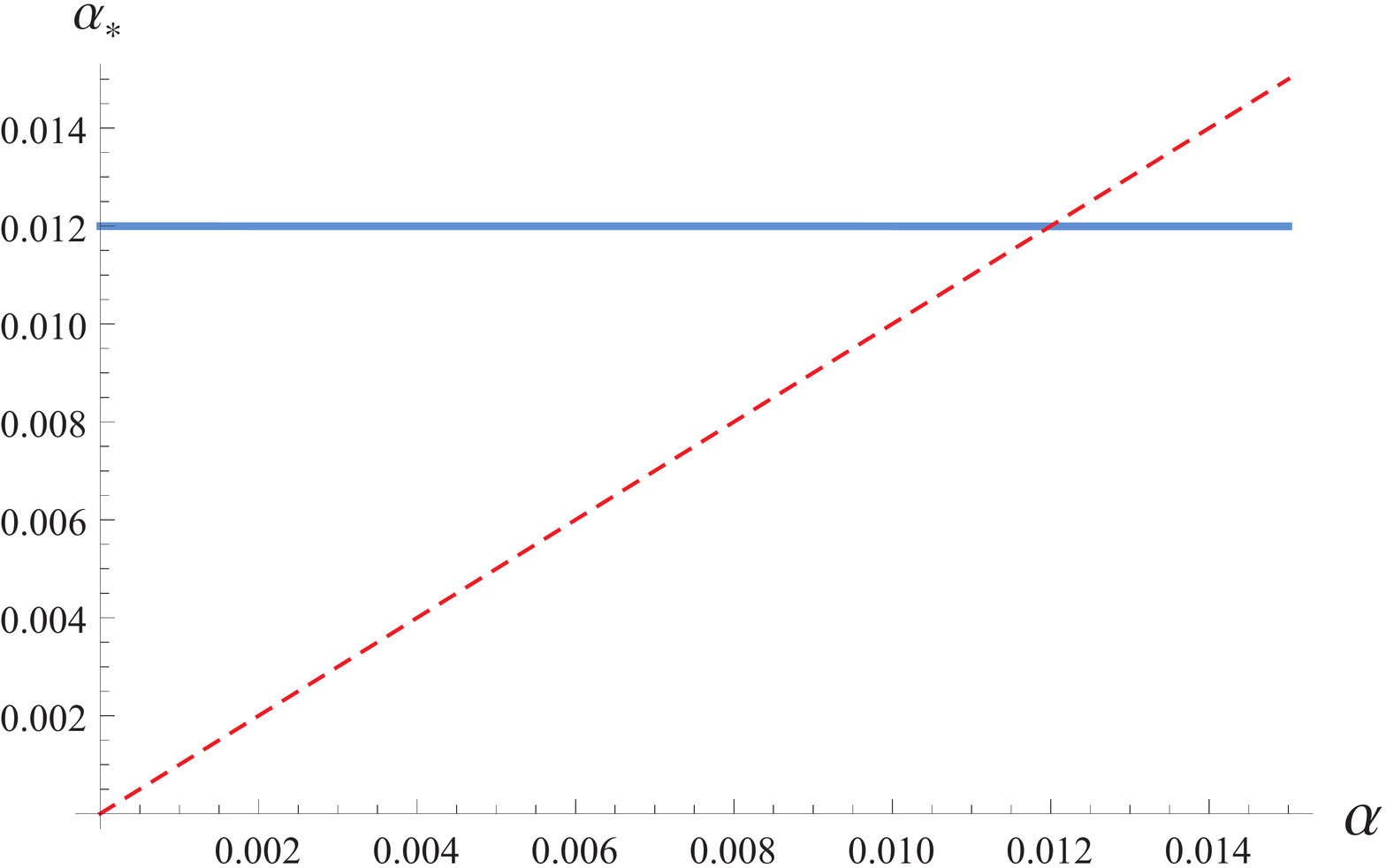}
 	
 		\caption{\footnotesize   
 			The determination of the  physical range of $\alpha$ using the  relation~(\ref{eq:constrain-alpha}), in scheme 2. (Here,  we have chosen $T_2\equiv T_H=60$, $V_2=33000$, $T_4\equiv T_{\rm C}=30$, $V_4=15500$,   which give the upper bound on  $\alpha$ as approximately 0.0119936.)}   \label{fig:scheme2GB-physicalRangeof_alpha}
 	}
 \end{figure}
 
 \begin{figure}[h!]
 	{\centering
 		\subfloat[]{\includegraphics[width=2.5in]{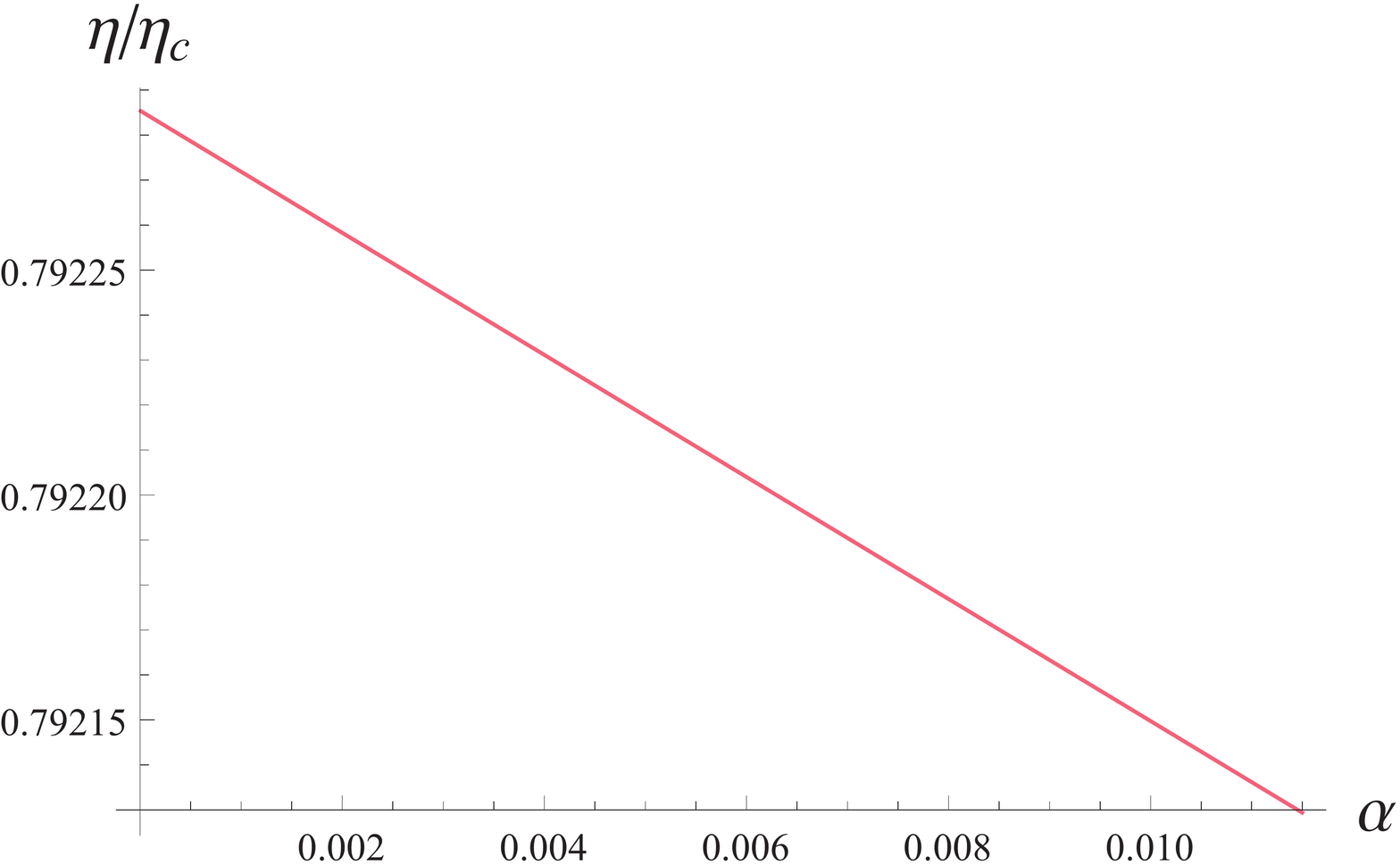} }\hspace{1.9cm}  	
 		\subfloat[]{\includegraphics[width=2.5in]{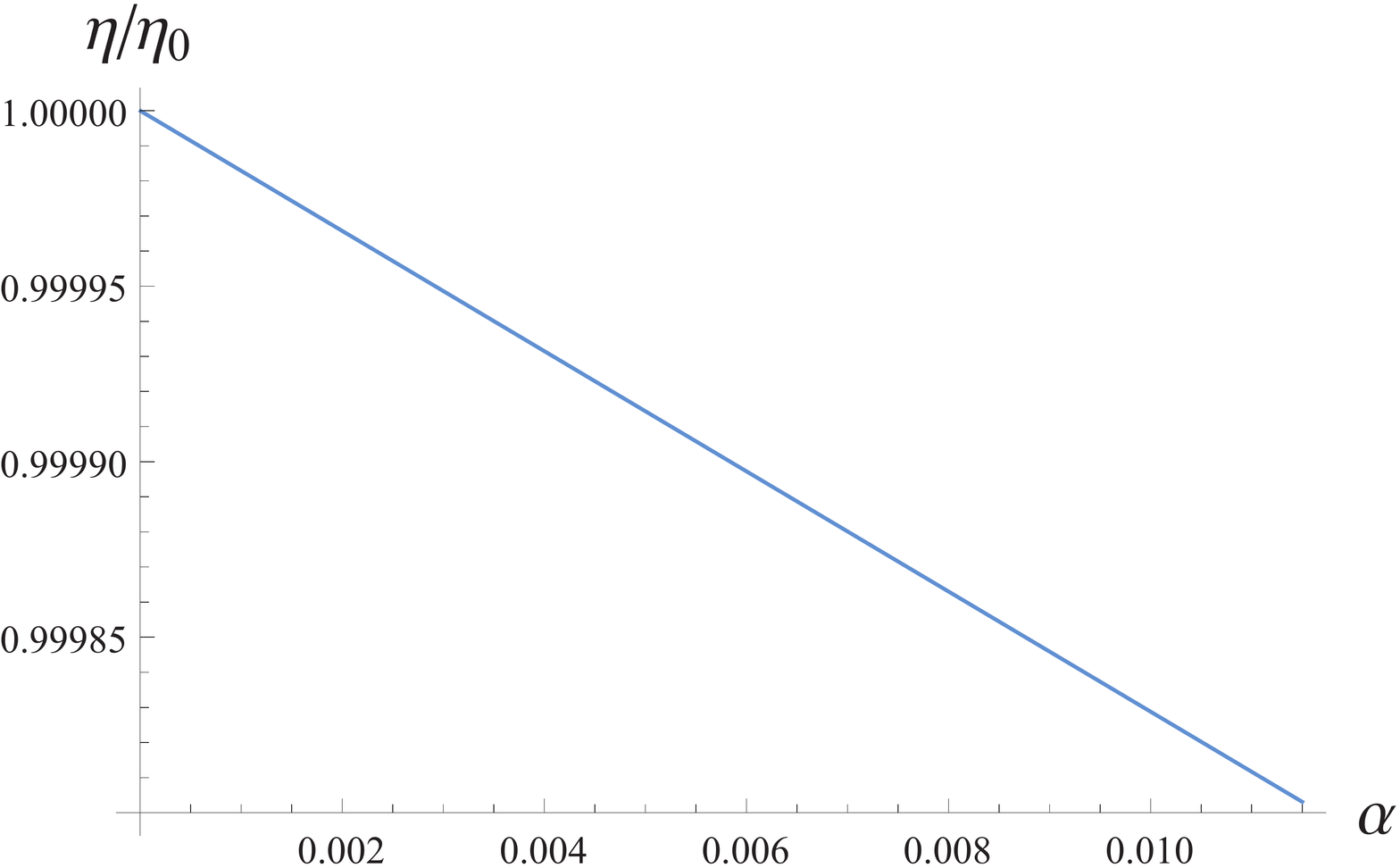} }
 		
 		\caption{\footnotesize  In scheme 2, over the  physical range of $\alpha$ constrained  by the relation~(\ref{eq:constrain-alpha}), (a) The ratio $\eta/\eta_{\rm C}$ {\it vs} $\alpha$, and  (b)  The ratio $\eta/\eta_0$  {\it vs} $\alpha$.   (See the  caption of fig (\ref{fig:scheme2GB-physicalRangeof_alpha}) for parameter values.)}   \label{fig:scheme1_compare2}
 	} 
 \end{figure}
\noindent
We note from figures (\ref{fig:scheme1_TC_and_itac}),(\ref{fig:scheme1_compare}),(\ref{fig:scheme2GB-physicalRangeof_alpha}) and (\ref{fig:scheme1_compare2}) that in GB black holes, the behavior of efficiency of heat engines is unaffected by the lack of charge parameter $q$ and results are identical to those found in~\cite{Johnson:2015ekr} where charged black holes were used as working substances.\\

\section{Critical Black Holes and Heat Engines}

\begin{figure}[h]
	\begin{center}
		\centering
				\includegraphics[width=3.2in]{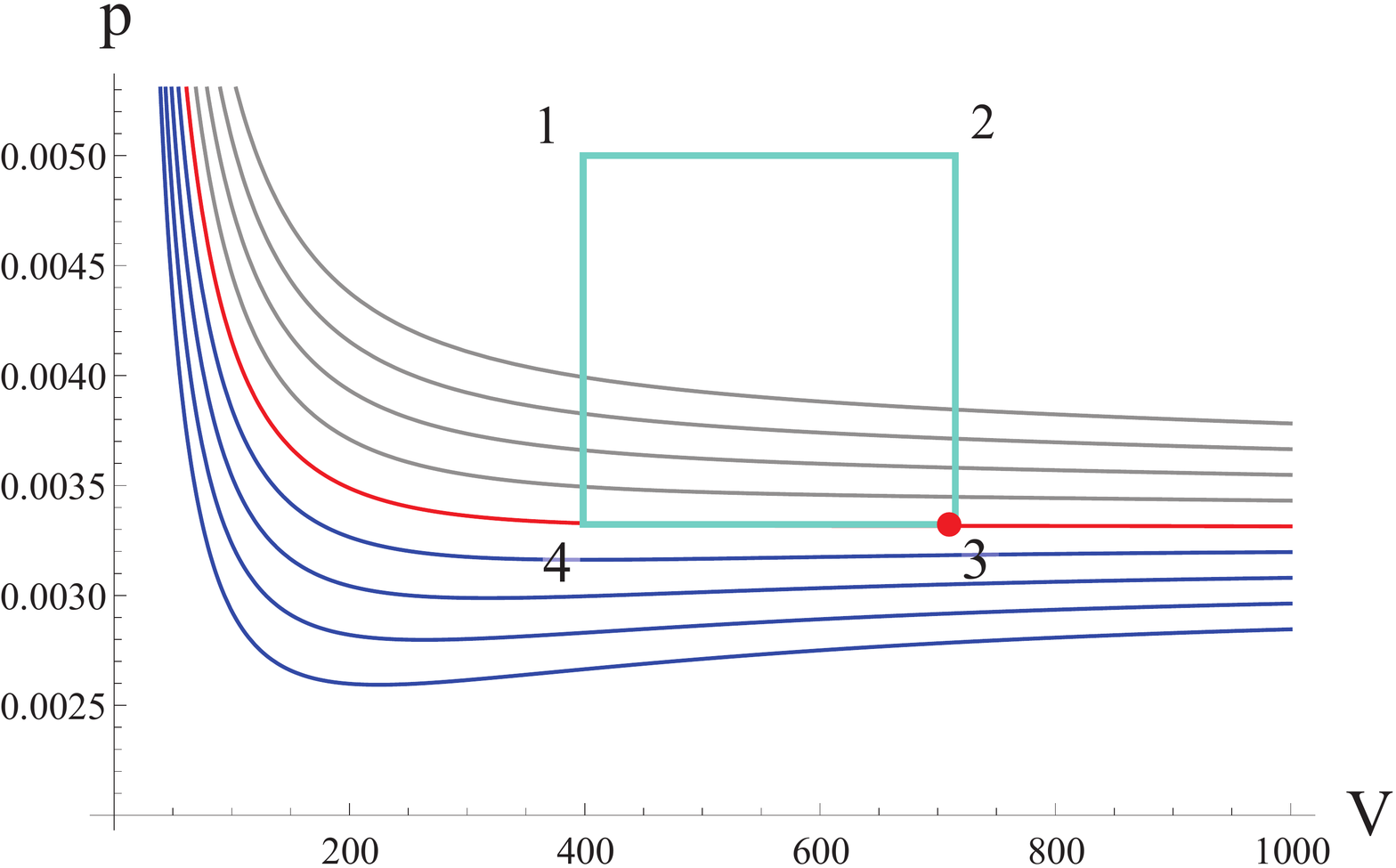}  
			
			\caption{ Sample isotherms for $\alpha = 2$ obeying the constraint ( \ref{eq:constrain-alpha}). The central (red) isotherm is for critical temperature $T_{cr}$, the gray colored isotherms are for $T > T_{cr}$ and the blue colored isotherms are for $T < T_{cr}$. The critical point is highlighted with red dot where  the corner 3 of the engine cycle is  placed.}   \label{fig:isotherms}
		
	\end{center}
\end{figure}

\noindent
Now, the critical region can be understood from the behaviour of the equation of state for different isotherms as seen from figure (\ref{fig:isotherms}). For a given $\alpha$, there exist a critical temperature $T_{cr}$,  below which the  equation of state exhibits  the first order phase transition between the small and large black holes which is reminiscent of the liquid/gas phase transition of van der Waals fluid. At  $ T= T_{cr}$, first order phase transitions terminate in a second order critical point.  \\

\noindent
In particular, in the $p - V$ plane, the point of inflection: $\partial p/\partial V=\partial^2 p /\partial V^2=0$  determines the critical point as~\cite{Cai:2013qga}:
\begin{equation}
\label{eq:critical}
\quad p_{\rm cr}=\frac{1}{48\pi \alpha}\ , \quad V_{\rm cr}= 18\pi^2 \alpha^2 \ ,   \quad T_{\rm cr}= \frac{1}{\pi\sqrt{24\alpha}}\ , \quad \text{where} \, \, \, r_{\rm cr}= \sqrt{6\alpha}.
\end{equation}

\noindent
Since for a given $\alpha$, $S$ and $V$ are not independent, the specific heat in a isochoric process vanishes, where as it is a non-vanishing quantity in an isobaric process~\cite{Johnson:2015ekr,Cai:2013qga}, i.e.,

\begin{equation}
	C_V=0 \ ; \, \, \,   C_p=\frac{3\pi^2}{2}\left(\frac{(8\pi pr_+^2+3)(r_+^2+2\alpha)^2r_+}{8\pi pr_+^2(r_+^2+6\alpha)-3r_+^2+6\alpha}\right)\ .
\end{equation}


\noindent
We define the engine cycle as a rectangle in $p-V$ plane (which is a natural choice when $C_V = 0$ ~\cite{Johnson:2014yja}), and the equation (\ref{eq:efficiency-prototype}) can be employed in computing the efficiency. It was stated in~\cite{PhysRevLett.114.050601,power_of_a_critical_heat} that, probing the engine containing the critical point (or even close to critical point) results in approaching the Carnot's efficiency having the finite power. A working example of this feature was constructed in~\cite{Johnson:2017hxu} in the context of charged-AdS black holes in the large charge limit. Following~\cite{Johnson:2017hxu}, we place the critical point at the corner 3 (see fig. \ref{fig:isotherms}) and choose the boundaries of the cycle\footnote{One can choose different boundaries and place the critical point at other corners, but we stick to the choice in~\cite{Johnson:2017hxu} for later comparison.} such that
\begin{eqnarray}
 p_3 &= &p_4 =p_{\rm cr}, \nonumber \\ 
  p_1 &= & p_2  =  3p_{\rm cr}/2,  \nonumber \\
    V_2 &=& V_3=V_{\rm cr},  \nonumber \\ \text{and} \, \, \, \,  V_1 &=& V_4= V_{\rm cr}-V_{\rm cr}L/\alpha,  \, \, \, \text{where  \, L is a constant with dimensions of $\alpha$.}
\end{eqnarray}  
 With this set up, the work done can be readily calculated simply as the area $\Delta p \Delta V$ of the cycle given by $W =   p_{\rm cr}V_{\rm cr}L/2\alpha=\frac{3\pi}{16}L$. This work is finite and independent of $\alpha$, and the heat inflow $Q_H$ is given by:
  \begin{eqnarray}
 Q_H & = & M_2 -M_1  \nonumber \\
 &=& \frac{9 \pi}{16} \Bigg(L + 4\alpha \bigg(1-\sqrt{\Big(1-\frac{L}{\alpha}\Big)}\bigg)\Bigg).
 \label{eq: Exact	QH } \end{eqnarray} \par 
 Equation (\ref{eq: Exact	QH }) above shows that  as $\alpha$  increases, $Q_H$ decreases. Therefore the efficiency $\eta$ increases with $\alpha$. This result drives us to consider the limit of large $\alpha$ ( similar to the limit of large charge $q$ in~\cite{Johnson:2017hxu}). In fact, the engine  is physical on raising $\alpha$ as pressures in the cycle obey the constraint (\ref{eq:constrain-alpha}), while temperature is positive at any $\alpha$. However, large $\alpha$ affects the cycle to reduce its   height (as $\Delta p \sim \alpha^{-1}$), while increases its width (as $\Delta V \sim \alpha$), so that work is finite at any $\alpha$. \\
 \noindent
 The large $\alpha$ expansion for inflow of heat $Q_H$ (eq. \ref{eq: Exact	QH }) reads as: 
\begin{equation}
Q_H= \frac{27\pi}{16}L+\frac{9\pi}{32}{\frac {{L}^{2}}{\alpha}}+\frac{9\pi}{64}{\frac {{L}^{3}}{\alpha^2}}+\frac {45\pi}{512}\frac{L^4}{\alpha^3}+\frac {63\pi}{1024}\frac{L^5}{\alpha^4}+O \left( {\alpha}^{-5} \right)\ ,
\end{equation}
where as the efficiency $\eta=W/Q_H$ in the limit of large $\alpha$ is:
\begin{equation}
\eta=\frac{1}{9}-\frac{1}{54}{\frac {L}{\alpha}}-\frac{1}{162}{\frac {{L}^{2}}{{\alpha}^{2}}}-\frac{25}{7776}{\frac {{L}^{3}}{{\alpha}^{3}}}-\frac{95}{46656}{\frac {{L}^{4}}{{\alpha}^{4}}}+O \left( {\alpha}^{-5} \right) \ .
\end{equation}

\begin{figure}[h]
	\begin{center}
		{\centering
		\subfloat[]	{\includegraphics[width=2.8in]{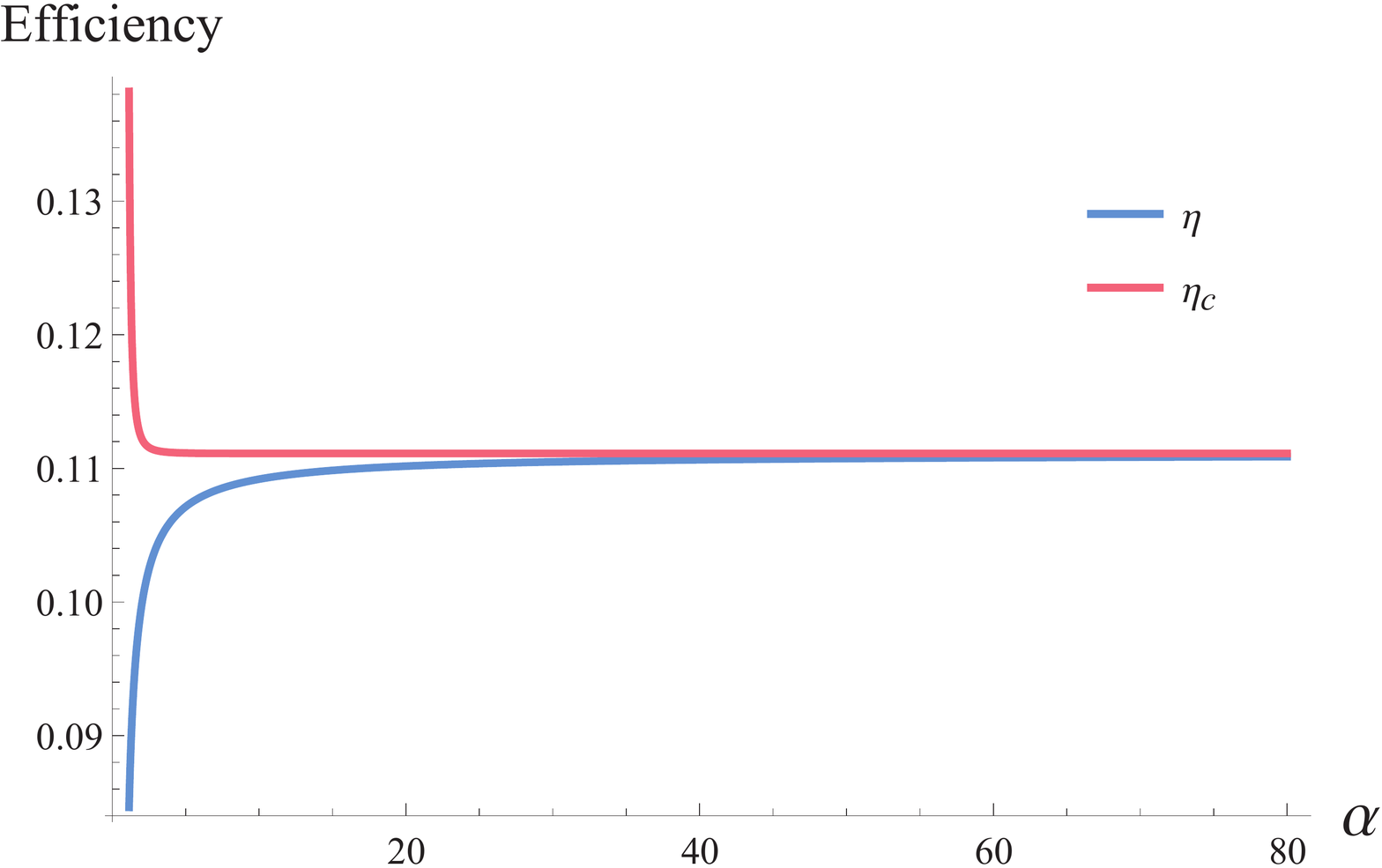} } \hspace {1.3cm}
	\subfloat[] 	{\includegraphics[width=2.55in]{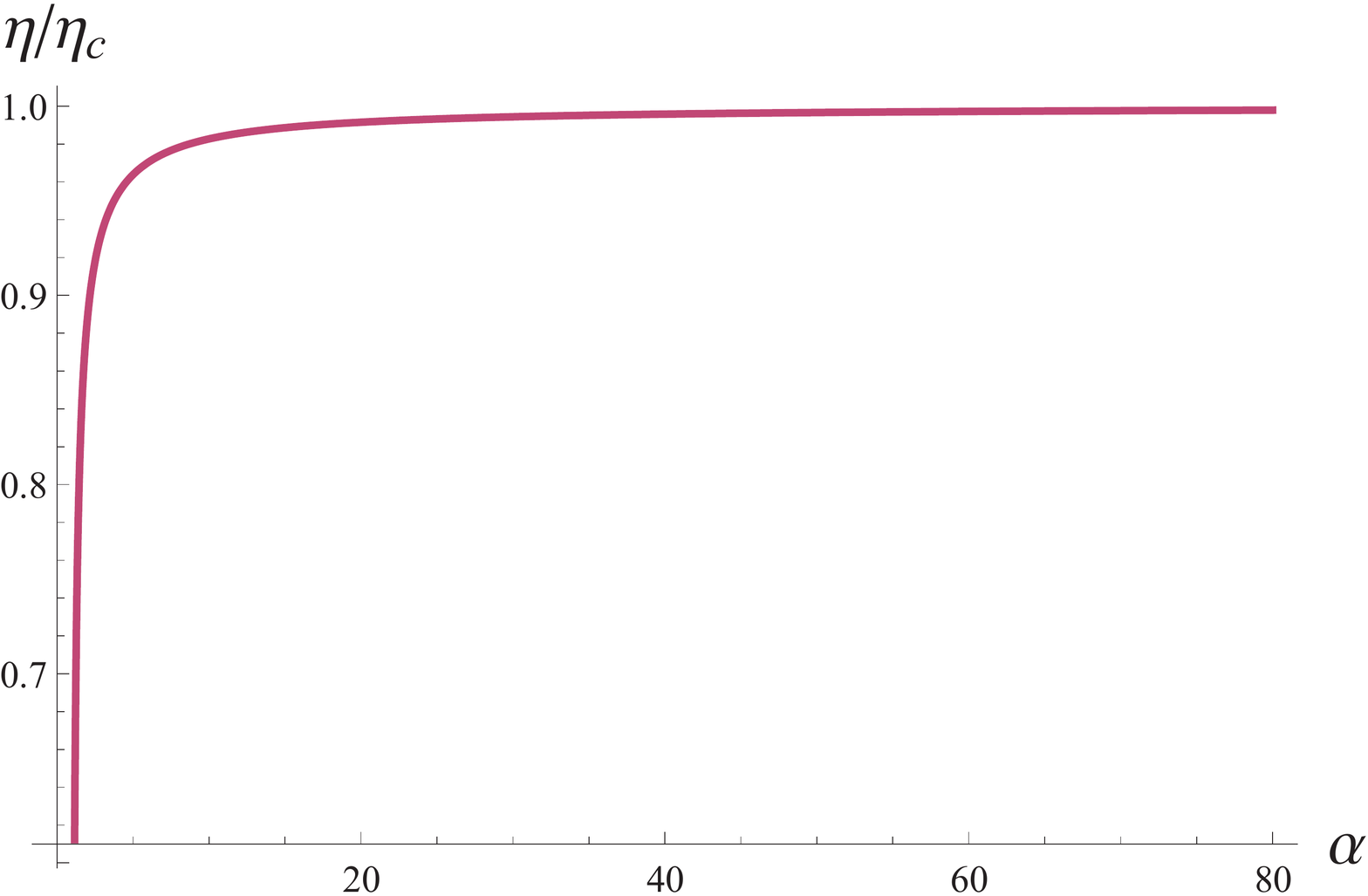} }
			\caption{ The behaviour of (a) $\eta, \, \eta_{\rm C}$   and (b) the ratio $\eta/\eta_{\rm C}$  with $\alpha$ ( Here, $L=1$ is used.)}   \label{fig:ita by itac}
		}
	\end{center}
\end{figure}

\noindent
The Carnot efficiency $\eta_{\rm C}^{\phantom{C}}$, which is independent of working substance depends only on the lowest  and highest  temperatures between which the engine runs. Our engine has the highest temperature $T_H$ at corner 2, while the lowest temperature $T_C$ is at corner 4.  Plugging the chosen values  for ($p_2,V_2$) and ($p_4,V_4$) into eq. (\ref{eq:temperature}) gives $T_H$ as: 
\begin{equation}
 T_H = \frac{9}{8\pi \sqrt{24\alpha}}\, . 
\end{equation}
\text{while  the large $\alpha$ expansion  for $T_C$  is:}
\begin{equation} 
 T_C  =  \frac{1}{2\pi \sqrt{6\alpha}} -{\frac {{L}^{3}}{512 \sqrt{6} \,\pi\, \alpha^{7/2}}}-{\frac {\sqrt{3}\,{L}^{4}}{1024 \sqrt{2} \,\pi\, \alpha^{9/2}}}+O \left({\alpha}^{-11/2} \right)\, .
\end{equation}
  These temperatures provide the Carnot's efficiency at large $\alpha$ as:
\begin{equation}
\label{eq:carnot_critical}
\eta_{\rm C}^{\phantom{C}}=1-\frac{T_C}{T_H} = {\frac{1}{9}}+\frac{1}{288}{\frac {{L}^{3}}{\alpha^{3}}}+\frac{1}{192}{\frac {L^{4}
	}{\alpha^{4}}}+O \left( {\alpha}^{-
	5} \right) 
\ .\end{equation}
Since, the GB parameter $\alpha$ seems to mimic the behavior of charge parameter $q$, we would now like to compare ours results with those of charged black holes  in $5$ dimensional AdS spacetime\footnote{For efficiency computations, particularly in D=4, see~\cite{Johnson:2017hxu}}. In this context, the expressions for temperature,  entropy,  thermodynamic volume and mass (enthalpy) of the black hole are respectively given by~\cite{Johnson:2017asf,Gunasekaran:2012dq}:
\begin{eqnarray}
T &=&\frac{1}{2\pi}\left(\frac{1}{r_+}-\frac{q^2}{r_+^5}+\frac{8\pi p}{3}r_+\right)\ , \label{RN temp eq} \\
S &=& \frac{\pi^2}{2}r_+^3 \ , \\
V &=&  \frac{\pi^2}{2}r_+^4 \ , \\
M (S, p) &=& \frac{3\pi}{8}\Big(\frac{2S}{\pi^2}\Big)^{\frac{-2}{3}} \Big\{ \Big(\frac{2S}{\pi^2}\Big)^{\frac{4}{3}} + q^2 + \frac{16pS^2}{3\pi^3}\Big\} \ . \label{RN enthalpy eq}
\end{eqnarray}
The equation of state $p (V , T)$ can be obtained from the eq. (\ref{RN temp eq}), as:
\begin{equation}
\label{RN EOS}
p= \frac{3}{8\sqrt{2V}}\Big(2\sqrt{\pi} T (2V)^{\frac{1}{4}} - 1 + \frac{\pi^2 q^2}{2V}\Big) \ .
\end{equation}
This equation of state shows the small/large black hole phase transitions in $p-V$ plane similar to 4-dimensional case, and there exists a first order phase transition line terminating at the second order critical point given by~\cite{Gunasekaran:2012dq}:
\begin{equation}
\label{eq:RN critical quantities}
\quad p_{\rm cr}=\frac{1}{4\sqrt{15}\pi q}\ , \quad V_{\rm cr}= \frac{15}{2}\pi^2 q^2 \ ,   \quad T_{\rm cr}= \frac{4}{5\pi (15)^{\frac{1}{4}}\sqrt{q}}\ , 
\end{equation}
where \, \,  $ r_{\rm cr}= (15) ^{\frac{1}{4}}\sqrt{q} $  \, \, and \, \,  $ S_{\rm cr} = \frac{\pi^2}{2}(15q^2)^{\frac{3}{4}}$.
The specific heats at isochoric and isobaric process are~\cite{Johnson:2015ekr}:
\begin{equation}
C_V=0 \ ; \, \, \,   C_p=\frac{3\pi^2}{2}r_+^3\left(\frac{8\pi pr_+^6 + 3r_+^4 - 3q^2}{8\pi pr_+^6 - 3r_+^4 + 15q^2}\right)\ .
\end{equation}
Similar to the Gauss-Bonnet case, the rectangular cycle defined as
\begin{eqnarray}
p_3 &= &p_4 =p_{\rm cr}, \nonumber \\ 
p_1 &= & p_2  =  3p_{\rm cr}/2,  \nonumber \\
V_2 &=& V_3=V_{\rm cr},  \nonumber \\ \text{and} \, \, \, \,  V_1 &=& V_4= V_{\rm cr}-V_{\rm cr}L/q,  
\end{eqnarray} 
produces the finite work $ W = \frac{\pi \sqrt{15}}{16}L$ at any $q$ for the inflow of heat $Q_H$:
\begin{eqnarray}
Q_H & = & M_2 -M_1  \nonumber \\
&=&  \sqrt{\frac{3}{5}} \, \pi \Bigg\{\frac{15L\big(2+\sqrt{1-\frac{L}{q}}\big)  - 32q\big(1-\sqrt{1-\frac{L}{q}}\big)}{16 \, \sqrt{1-\frac{L}{q}}}\Bigg\} \ . \end{eqnarray}
Its large $q$ expansion is:
\begin{equation}
Q_H= \frac{29\pi}{16}\sqrt{\frac{3}{5}} \, L+\frac{3\pi}{16}\sqrt{\frac{3}{5}}\, {\frac {{L}^{2}}{q}}+\frac{\sqrt{15} \, \pi}{64}{\frac {{L}^{3}}{q^2}}+\frac {\sqrt{15}\, \pi}{128}\frac{L^4}{q^3}+\frac {21\pi}{1024}\sqrt{\frac{3}{5}}\, \frac{L^5}{q^4}+O \left( {q}^{-5} \right)\ ,
\end{equation}
while the efficiency $\eta =W/Q_H$ at large $q$ takes the form as:
\begin{equation}
\eta=\frac{5}{29}-\frac{15}{841}{\frac {L}{q}}-\frac{545}{97556}{\frac {{L}^{2}}{{q}^{2}}}-\frac{13405}{5658248}{\frac {{L}^{3}}{{q}^{3}}}-\frac{1418425}{1312713536}{\frac {{L}^{4}}{{q}^{4}}}+O \left( {q}^{-5} \right) \ .
\end{equation} 
Our engine has the highest temperature $T_H$ at corner 2, while the lowest temperature $T_C$ is at corner 4.  Plugging the chosen values  for ($p_2,V_2$) and ($p_4,V_4$) into eq. (\ref{RN temp eq}) gives, 
\begin{equation}
T_H = \frac{29}{2\pi (15)^{\frac{5}{4}} \, \sqrt{q}}\, . 
\end{equation}
\text{while  the large $q$ expansion  for $T_C$  is:}
\begin{equation} 
T_C  = \frac{4}{5 \pi (15)^{\frac{1}{4}}} \, \frac{1}{q^{1/2}} -\frac{1}{96 \pi (15)^{\frac{1}{4}}} \, \frac{L^3}{q^{7/2}} -\frac{29}{1536 \pi (15)^{\frac{1}{4}}} \, \frac{L^4}{q^{9/2}} +O \left({q}^{-11/2} \right)\, .
\end{equation}
These temperatures provide the Carnot's efficiency as:
\begin{equation}
\label{eq:carnot_critical}
\eta_{\rm C}^{\phantom{C}}=1-\frac{T_C}{T_H} = {\frac{5}{29}}+\frac{5}{464}{\frac {{L}^{3}}{q^{3}}}+\frac{5}{256}{\frac {L^{4}
	}{q^{4}}} +\frac{393}{14848}{\frac {{L}^{5}}{q^{5}}}+O \left( {q}^{-
	6} \right) 
\ .\end{equation}

\begin{figure}[h]
	\begin{center}
		{\centering
			\subfloat[]	{\includegraphics[width=2.6in]{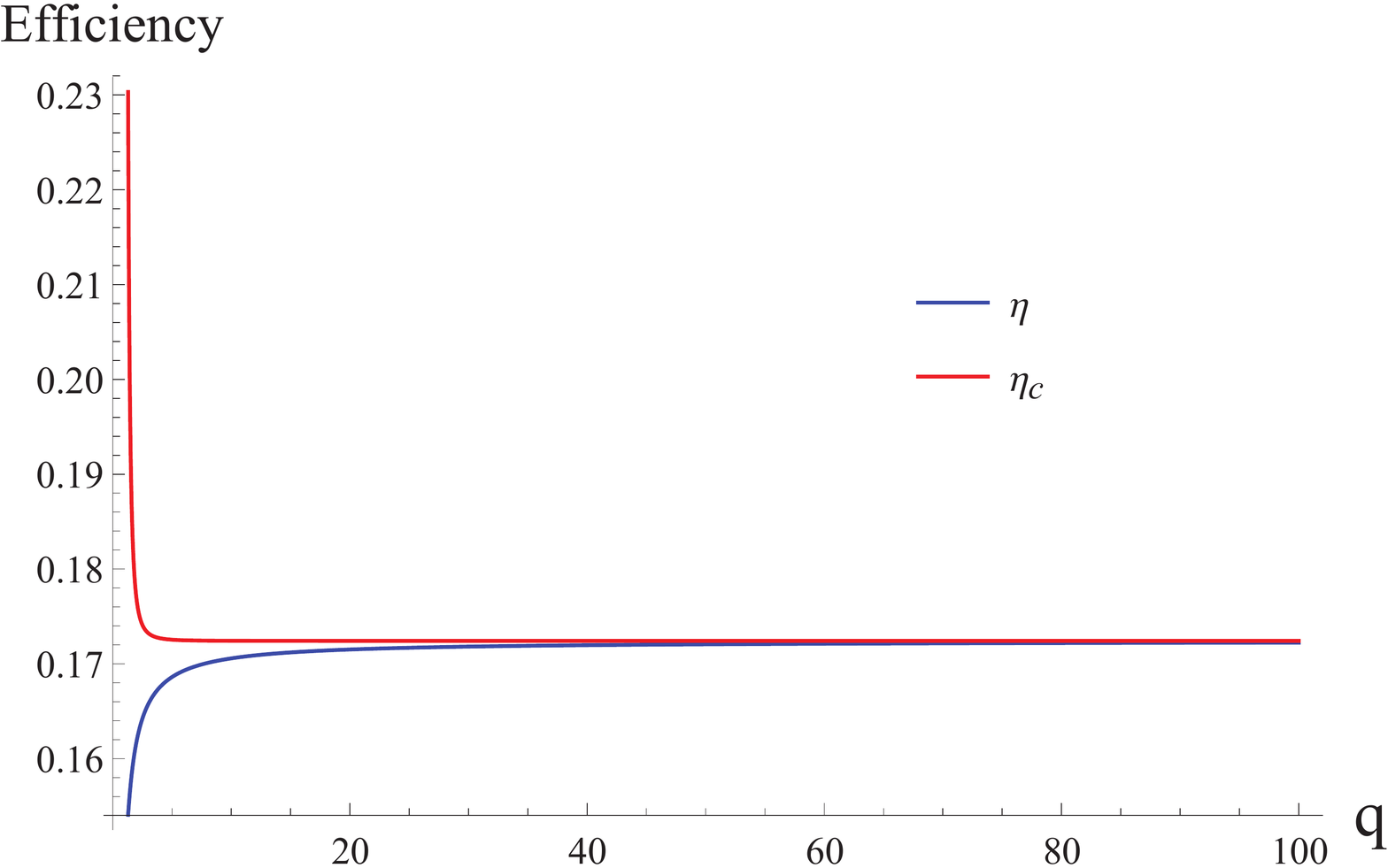} } \hspace {1.3cm}
			\subfloat[] 	{\includegraphics[width=2.45in]{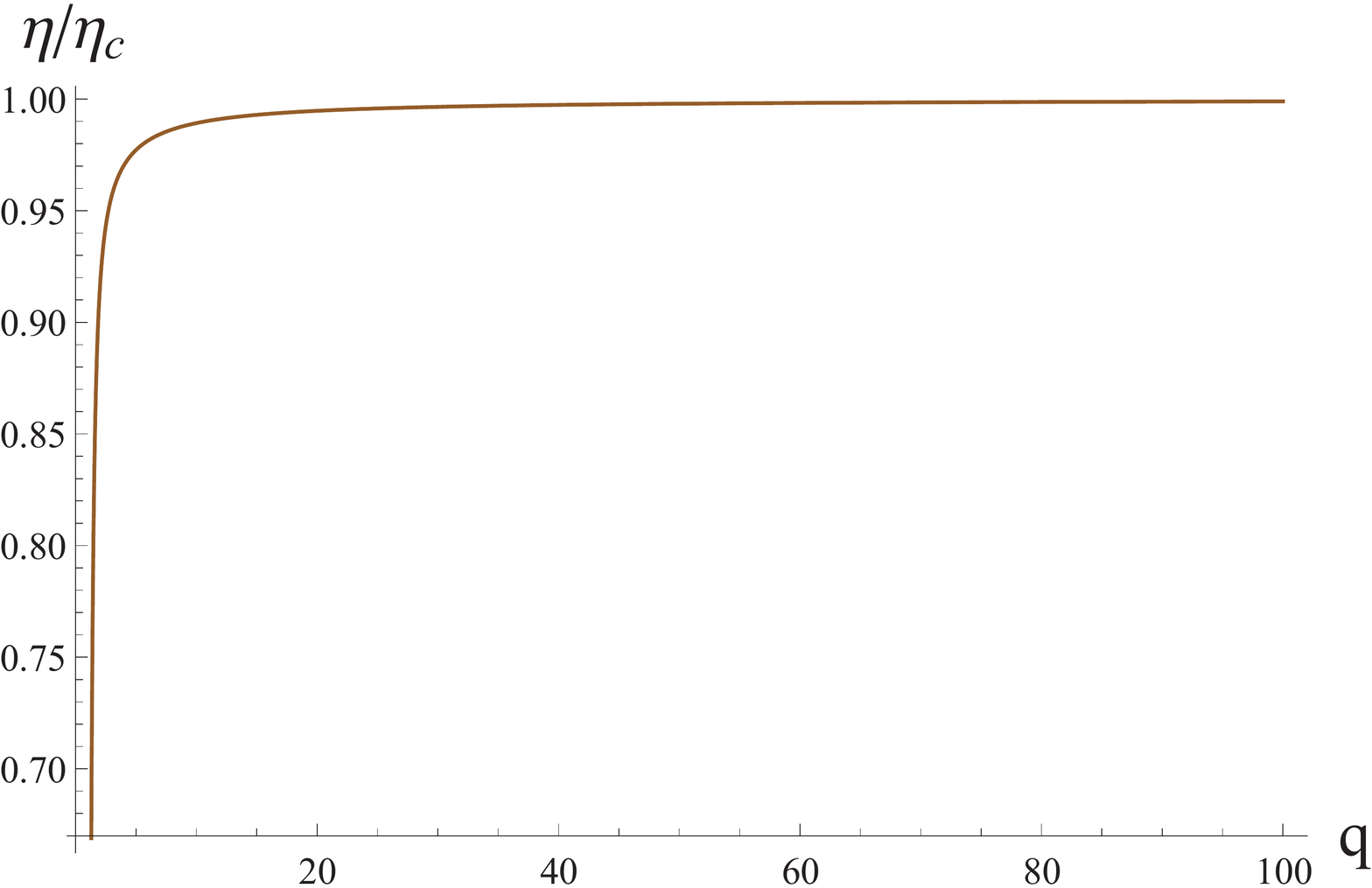} }
			\caption{ The behaviour of (a) $\eta, \, \eta_{\rm C}$   and (b) the ratio $\eta/\eta_{\rm C}$  with $q$ ( Here, $L=1$ is used.)}   \label{fig:RN ita by itac}
		}
	\end{center}
\end{figure}
From figures ( \ref{fig:ita by itac} \& \ref{fig:RN ita by itac}), we see that, in both the cases,  $\eta \rightarrow \eta_{\rm C}$ at large parameter values  with finite work. In fact, $\eta = \eta_{\rm C}$ is possible only when the respective parameter ($\alpha$ or $q$, depending on the case) go towards $\infty$, while the power vanishes in this limit, according to the \textit{universal trade off relation} between power and efficiency given by~\cite{PhysRevLett.117.190601,eff_vs_speed}:

\begin{equation}
\frac{W}{\tau}\leq{\bar\Theta} \frac{\eta(\eta_{\rm C}^{\phantom{C}}-\eta)}{T_C}\ , \label{trade off relation}
\end{equation}
where ${\bar\Theta}$ is a model dependent constant  of the engine.  
The right hand side quantity  in eqn.(\ref{trade off relation}) (divided by ${\bar\Theta}$) has the large $\alpha$ expansion for Gauss-Bonnet case as:

 \begin{equation} 
 \frac{\eta(\eta_{\rm C}-\eta)}{T_C}  =  \frac{2 \pi}{81 \sqrt{6}} \frac{L}{\alpha^{1/2}} +\frac{\pi}{243 \sqrt{6}} \frac{L^2}{\alpha^{3/2}}+ \frac{\pi}{162\sqrt{6}}\frac{L^3}{\alpha^{5/2}} + \frac{1987 \pi}{279936\sqrt{6}}\frac{L^4}{\alpha^{7/2}} + O \left({\alpha}^{-9/2}\right) \, ,
 \end{equation}
where as  the large $q$ expansion for charged black hole case is:
\begin{equation} 
\frac{\eta(\eta_{\rm C}-\eta)}{T_C}  =  \frac{375 \pi \, (15)^{\frac{1}{4}}}{97556} \frac{L}{q^{1/2}} +\frac{9125\pi \, (15)^{\frac{1}{4}}}{11316496} \frac{L^2}{q^{3/2}}+ \frac{3391875 \pi \, (15)^{\frac{1}{4}}}{1312713536}\frac{L^3}{q^{5/2}}+ O \left({q}^{-7/2}\right).
\end{equation}
The time $\tau$ taken to complete the cycle scales as $\tau \sim \alpha$ in Gauss-Bonnet case, where as $\tau \sim q$ in charged black hole case basing on the behaviour of critical pressures~\cite{Johnson:2017hxu}. Therefore, approaching  $\eta$ to $\eta_{\rm C}$ at large parameter values is carried out at finite power in Gauss-Bonnet black holes as well as in charged black holes.
\section{Remarks}
\begin{figure}[h]
	\begin{center}
		{\centering
			\subfloat[]	{\includegraphics[width=2.6in]{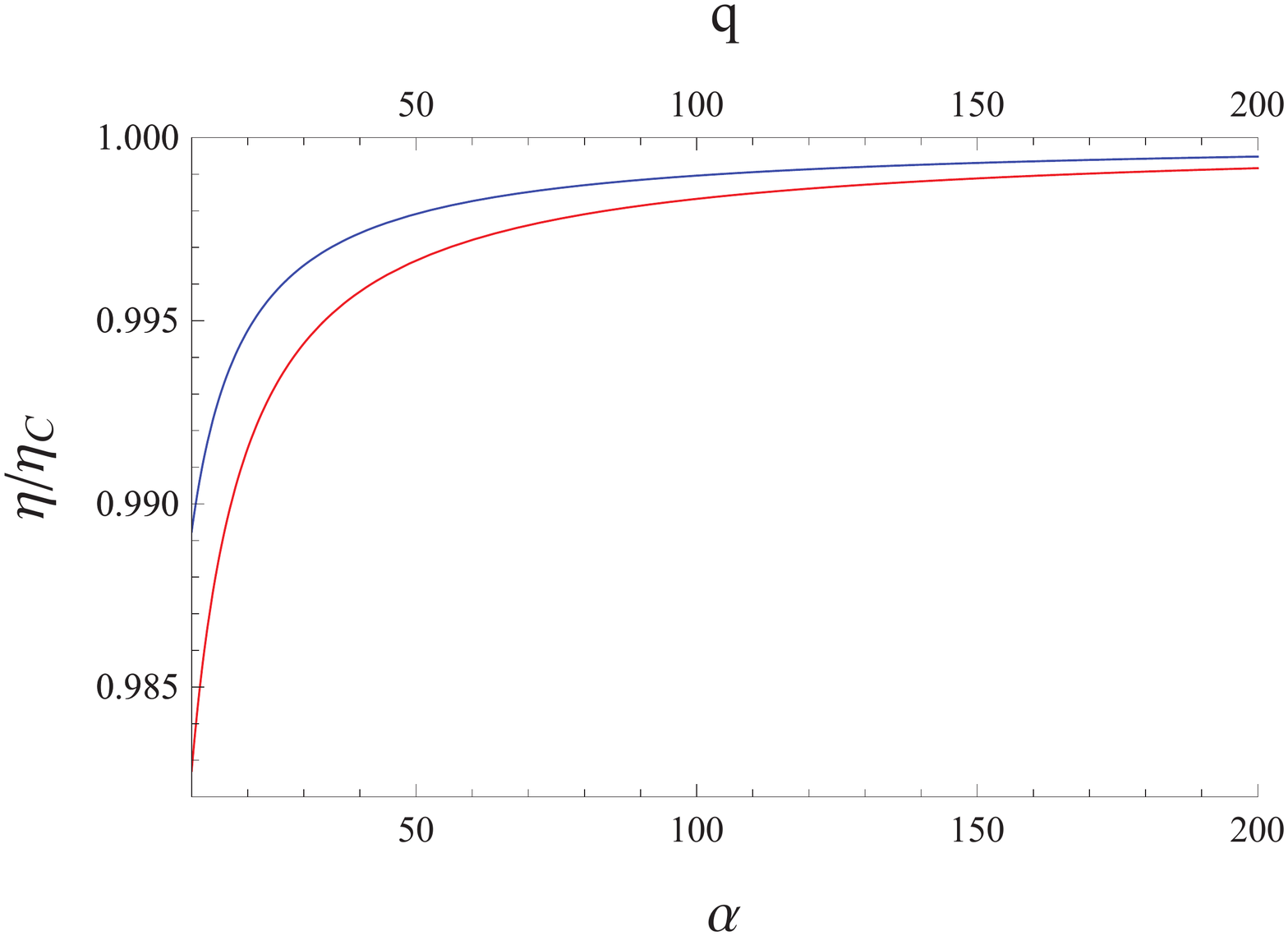} } \hspace {1.3cm}
			\subfloat[] 	{\includegraphics[width=2.45in]{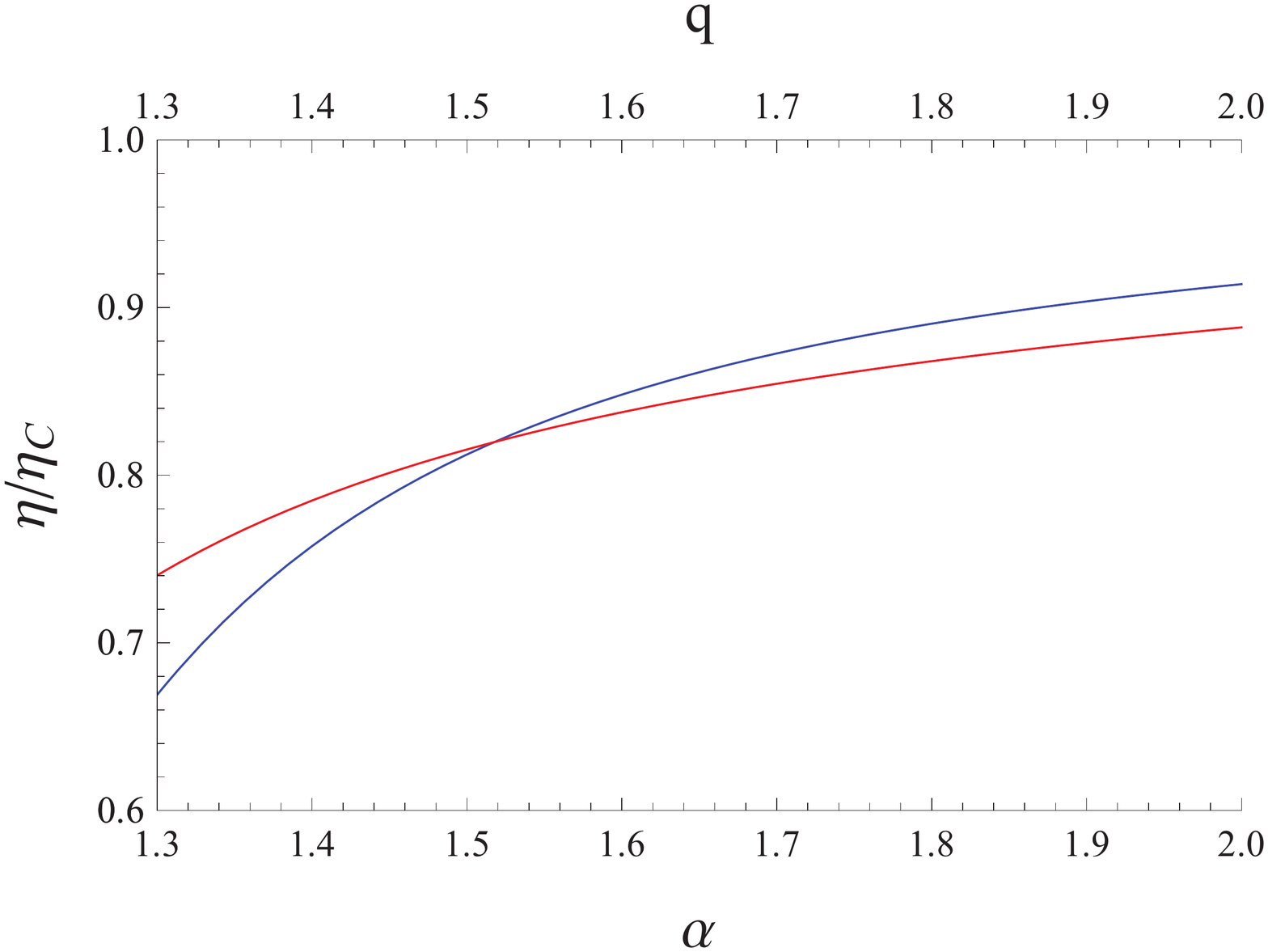} }
			\caption{ A comparative plot of the ratio $\eta/\eta_{\rm C}$  between charged black hole (Blue curve) and Gauss-Bonnet black hole (Red curve) ( we fix $L=1$) (a) Full range of $q$ and $\alpha$ and (b) Lower values of $q$ and $\alpha$}   \label{fig:RN_vs_GB}
		}
	\end{center}
\end{figure}

On comparison, figure (\ref{fig:RN_vs_GB}) shows that the approach to Carnot's efficiency at finite power at \textit{large} parameter values is faster in the case of charged black holes than in Gauss-Bonnet black holes. However, at smaller values of parameters, approach of  $\eta$ to $\eta_{\rm C}$ for the engine in case of the latter dominates over the former.  Let us also note that  $\eta$ and $\eta_{\rm C}$ converge to 5/29 for 5D charged black hole and 1/9 for 5D Gauss-Bonnet black hole, while for the 4D charged black hole they converge to 3/19~\cite{Johnson:2017hxu}. The convergent point of  $\eta$ and $\eta_{\rm C}$ for the engine \textit{defined as above}, seem to follow the relation:
\begin{equation}
\text{Convergent point of} \, \, \eta \, \, \text{and} \, \, \eta_{\rm C} = \frac{\rho_{cr}}{2 + \rho_{cr}} \, ,
\end{equation} 
where $\rho_{cr}$ is the critical ratio~\cite{Gunasekaran:2012dq,Cai:2013qga}:
\begin{equation*}
\rho_{cr} = \Bigg\{
\begin{array}{ccc}
3/8, & \text{4D charged black hole} \\
5/12, & \text{5D charged black hole}\\
1/4, & \text{5D Gauss-Bonnet black hole} \, .	
\end{array}
\end{equation*}
These features of heat engines at criticality for 5D GB black holes discussed in comparison to 5D charged black holes are counter intuitive and their holographic implications are worth understanding. 

\begin{figure}[h!]
	{\centering
		\includegraphics[width=2.8in]{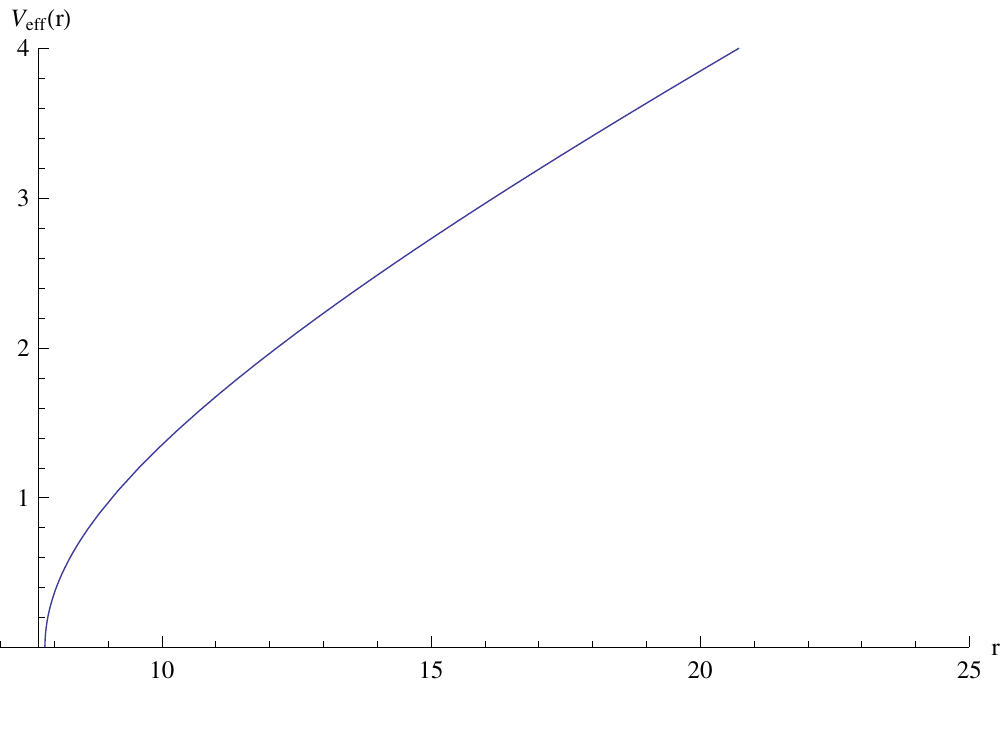}
		
		\caption{\footnotesize   
			Effective Potential for $L=0$, $\alpha=10$, $\mu=1$.}   \label{fig:eff}
	}
\end{figure}

\noindent
We can now take a closer look at the critical region by studying the metric function of Gauss-Bonnet black hole with critical values inserted, i.e.,
\begin{equation}
Y_{\rm cr}(r)=1+\frac{r^2}{2\alpha}\left(1-\sqrt{1+\frac{4\,\alpha \, m_{\rm cr}}{r^{4}}-\frac{4\alpha}{l_{\rm cr}^2}}\right) \ ,
\label{eq:Ycrit}
\end{equation}
\begin{equation}
m_{\rm cr} = 16\,\alpha , \quad l^2_{\rm cr} =  72\, \alpha
\end{equation}
Following the idea of a  coupled system leading to Carnot efficiency at criticality ~\cite{Johnson:2017hxu,PhysRevLett.114.050601,power_of_a_critical_heat}, it is worth studying
the picture of $\alpha$ interacting constituent objects. Consider, a particle of mass $\mu$ moving in the background of this critical black hole in the probe approximation. Following the methods in~\cite{Johnson:2017asf,Chandrasekhar1984,doi:10.1142/S0217732311037261}, the effective potential is seen to be
\begin{equation}
V_{\rm eff}(r) = \sqrt{Y_{\rm cr}(r)}\sqrt{\mu^2+\frac{L^2}{r^2}}\ ,
\end{equation}
where $L$ is the angular momentum of the particle. This is plotted in figure (\ref{fig:eff}) showing an attractive and binding behavior, though there is no local minimum (unlike the case of a probe charged particle~\cite{Johnson:2017asf}).

In fact, one can study the critical Gauss-Bonnet black hole in the double limit, where the parameter $\alpha$ is taken to be large while at the same time nearing the horizon. That is, one writes~\cite{Johnson:2017asf}, $r=r_++\epsilon\sigma$ and $t= \tau/\epsilon$, where, $Y(r=r_+)=0$ and $Y'(r=r_+) = 4\pi T_{\rm cr}$.
The near horizon limit is obtained by taking $\epsilon\to0$ while at the same time taking the large $\alpha$ limit by holding  $\epsilon \sqrt{\alpha}$ fixed. The metric in (\ref{eq:staticform}) goes over to $ds^2 = -{(4\pi {\widetilde T}_{\rm cr})\,\,\sigma}d\tau^2+\frac{1}{(4\pi {\widetilde T}_{\rm cr})}\frac{d\sigma^2}{\sigma }+d{\mathbb R}^{3}$. Here, ${\widetilde T}_{\rm cr}$ is $T_{\rm cr}$ in equation~(\ref{eq:critical}) with $\sqrt{\alpha}$ replaced by~${\epsilon \sqrt{\alpha}}$. Also,  $\Lambda=0$  and  since the $S^{3}$ has infinite radius ($r_{\rm cr}$ diverges at large $\alpha$ from eqn. (\ref{eq:critical}), the metric there is essentially flat $d{\mathbb R}^{3}=dx_1^2+dx_2^2+dx_{3}^2\ $. Thus, this double limit results in a completely decoupled Rindler space-time with zero cosmological constant, exactly analogous to the one uncovered for critical charged black holes in~\cite{Johnson:2017asf}.


\bibliographystyle{apsrev4-1}

\bibliography{GBcritical}

\end{document}